\documentclass[10pt]{article}
\usepackage{amsmath, bm}
\usepackage{amssymb}
\usepackage{graphicx}
\usepackage{fancyhdr}

\def\cmda{Celest. Mech. Dyn. Astr. }

\def\jgr{J. Geophys. Res. }
\def\mnras{MNRAS }

\begin{document}

\setcounter{figure}{0}
\setcounter{table}{0}
\setcounter{footnote}{0}
\setcounter{equation}{0}

\begin{center}
{\it Submitted to the Proc. Journ\'ees 2014 ``Syst\`emes de r\'ef\'erence spatio-temporels'', 20 January 2015}
\end{center}

\vspace*{0.4cm}

\noindent {\Large ON THE DEFINITION AND USE OF THE ECLIPTIC IN MODERN}\\[0.08cm] 
\noindent {\Large ASTRONOMY}
\vspace*{0.7cm}

\noindent\hspace*{1.5cm} N. CAPITAINE$^1$, M. SOFFEL$^2$

\smallskip
\noindent\hspace*{1.5cm} $^1$SYRTE, Observatoire de Paris, CNRS, UPMC, \\
\noindent\hspace*{1.5cm} 61, avenue de l'Observatoire, 75014 -- Paris, France \\
\noindent\hspace*{1.5cm} e-mail: n.capitaine@obspm.fr\\[0.04cm] 
\noindent\hspace*{1.5cm} $^2$Lohrmann Observatory,\\
\noindent\hspace*{1.5cm} Dresden Technical University, 01062 Dresden, Germany\\
\noindent\hspace*{1.5cm} e-mail: michael.soffel@mailbox.tu-dresden.de

\vspace*{0.5cm}

\noindent {\large ABSTRACT.}  The ecliptic was a fundamental reference plane for astronomy from antiquity to the realization and use of the FK5 reference system. The situation has changed considerably with the adoption of the International Celestial Reference system (ICRS) by the IAU in 1998 and the IAU resolutions on reference systems that were adopted from 2000 to 2009. First, the ICRS has the property of being independent of epoch, ecliptic or equator. Second, the IAU~2000 resolutions, which specified the systems of space-time coordinates within the framework of General Relativity, for the solar system (the Barycentric Celestial Reference System, BCRS) and the Earth (the Geocentric Celestial Reference System, GCRS), did not refer to any ecliptic and did not provide a definition of a GCRS ecliptic. These resolutions also provided the definition of the pole of the nominal rotation axis (the Celestial intermediate pole, CIP) and of new origins on the equator (the Celestial and Terrestrial intermediate origins, CIO and TIO), which do not require the use of an ecliptic. Moreover, the models and standards adopted by the IAU~2006 and IAU~2009 resolutions are largely referred to the ICRS, BCRS, GCRS as well as to the new pole and origins. Therefore, the ecliptic has lost much of its importance. We review the consequences of these changes and improvements in the definition and use of the ecliptic and we discuss whether the concept of an ecliptic is still needed for some specific use in modern astronomy.
\vspace*{1cm}

\noindent {\large 1. INTRODUCTION} 
\smallskip

The ecliptic was a fundamental reference plane for astronomy (astrometry, solar system dynamics and measurements) from antiquity unto the realization and use of the FK5 reference system. This plane has been chosen because the equinox has historically provided a convenient fiducial point in the observation of the heavens and the passage of time. The situation has changed considerably with the adoption of the International Celestial Reference system (ICRS) by the IAU since 1998 and with the IAU resolutions on reference systems that were adopted between 2000 and 2009. These correspond to major improvements in concepts and realizations of astronomical reference systems, in the use of observational data and the accuracy of the models for the motions of the solar system objects and Earth's rotation\footnote[1]{The nomenclature associated with the new concepts and models has been provided by the IAU Working Group on ``Nomenclature for Fundamental Astronomy" (Capitaine et al.~2007; {\tt http://syrte.obspm.fr/iauWGnfa}).}.
In that modern context, which is consistent with General relativity (GR), the ecliptic is no more a fundamental plane and the concept of an ecliptic is not as clear as those of the other modern astronomical references. 

It is therefore necessary to review the consequences of these changes and improvements in the definition and use of the ecliptic and to discuss whether the concept of an ecliptic is still needed for some specific use in modern astronomy and whether a definition of the ecliptic in the GR framework is needed. 
\vspace*{0.7cm}

\noindent {\large 2. THE ASTRONOMICAL REFERENCE SYSTEMS AND RELATED PARAMETERS} 

\noindent {\it The IAU astronomical reference systems}

The International Celestial Reference System (ICRS) based on Very Long Baseline Interferometry (VLBI) observations of extragalactic radiosources has been adopted by the International Astronomical Union (IAU) since 1st January 1998 (IAU 1997 Resolution B2). The ICRS and the corresponding frame, the International Celestial Reference Frame (ICRF), replaced the FK5 system and the fundamental catalogue of stars FK5 (based on the determination of the ecliptic, the equator and the equinox), the Hipparcos catalogue being adopted as the primary realization of the ICRS in optical wavelengths. According to its definition, the ICRS is kinematically non-rotating with respect to the ensemble of distant extragalactic objects. It has no intrinsic orientation but was aligned close to the mean equator and dynamical equinox of J2000.0 for continuity with previous fundamental reference systems. Its orientation is independent of epoch, ecliptic or equator and is realized by a list of adopted coordinates of extragalactic sources. The current best realization of the ICRS is the second version of the ICRF, called ICRF2 (IAU~2009 Resolution~B2), to which a number of catalogues of celestial objects have been linked in order to densify it and make it accessible to astronomical observations at different wavelengths. This provides an idealized (quasi-inertial) barycentric coordinate system for measuring the positions and angular motions of the celestial objects, which is totally independent of the ecliptic.

The IAU~2000 resolutions specified the systems of space-time coordinates for the solar system and the Earth within the framework of General Relativity and provided clear procedures for the transformation between them. The Barycentric Celestial Reference System (BCRS), with its origin at the solar system barycenter and its axes oriented to match the ICRS (IAU~2006 Resolution B2) can be considered as being inertial if neglecting external galactic and extragalactic matter. It is  is used for solar system ephemerides, for interplanetary spacecraft navigation, for defining the positions of remote and moving objects, etc. The Geocentric Celestial Reference System (GCRS), with its origin at the center of mass of the Earth, might be called quasi-inertial, since the spatial axes are kinematically non-rotating with respect to the spatial BCRS-axes, whereas the geocenter is accelerated (Soffel et al. 2003). It is employed for the description of physical processes in the vicinity of Earth, for satellite theory, the dynamics of Earth (including Earth's rotation), as well as for the introduction of concepts such as the equator and the International Terrestrial Reference System (ITRS), etc.  
Such concepts and definitions do not refer to an ecliptic and do not provide a definition of a GCRS ecliptic (e.g. the J2000 GCRS ecliptic of Fig.~1 is not defined precisely).
\smallskip

\noindent {\it The Earth orientation parameters}

\begin{figure}[h!]
\begin{center}
\resizebox{4.5cm}{!}{\includegraphics{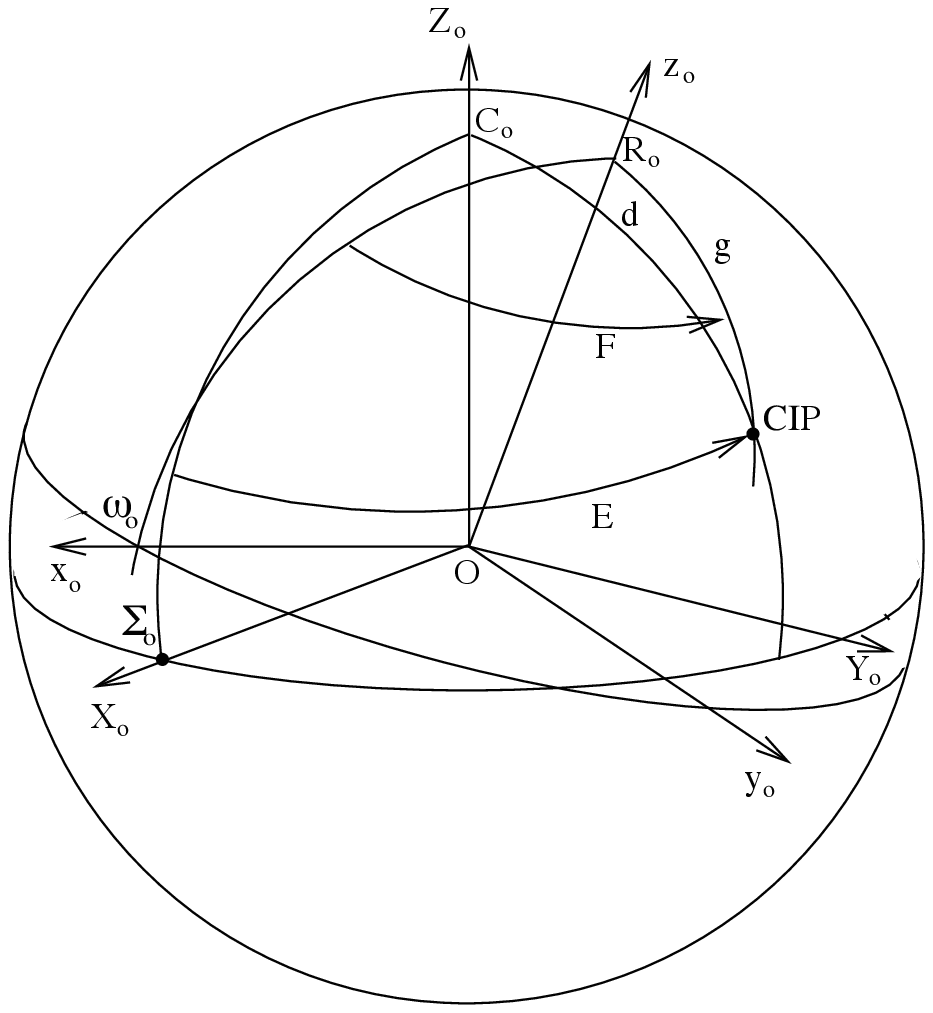} }
\hspace{1.5cm}
\resizebox{9.5cm}{!}{\includegraphics{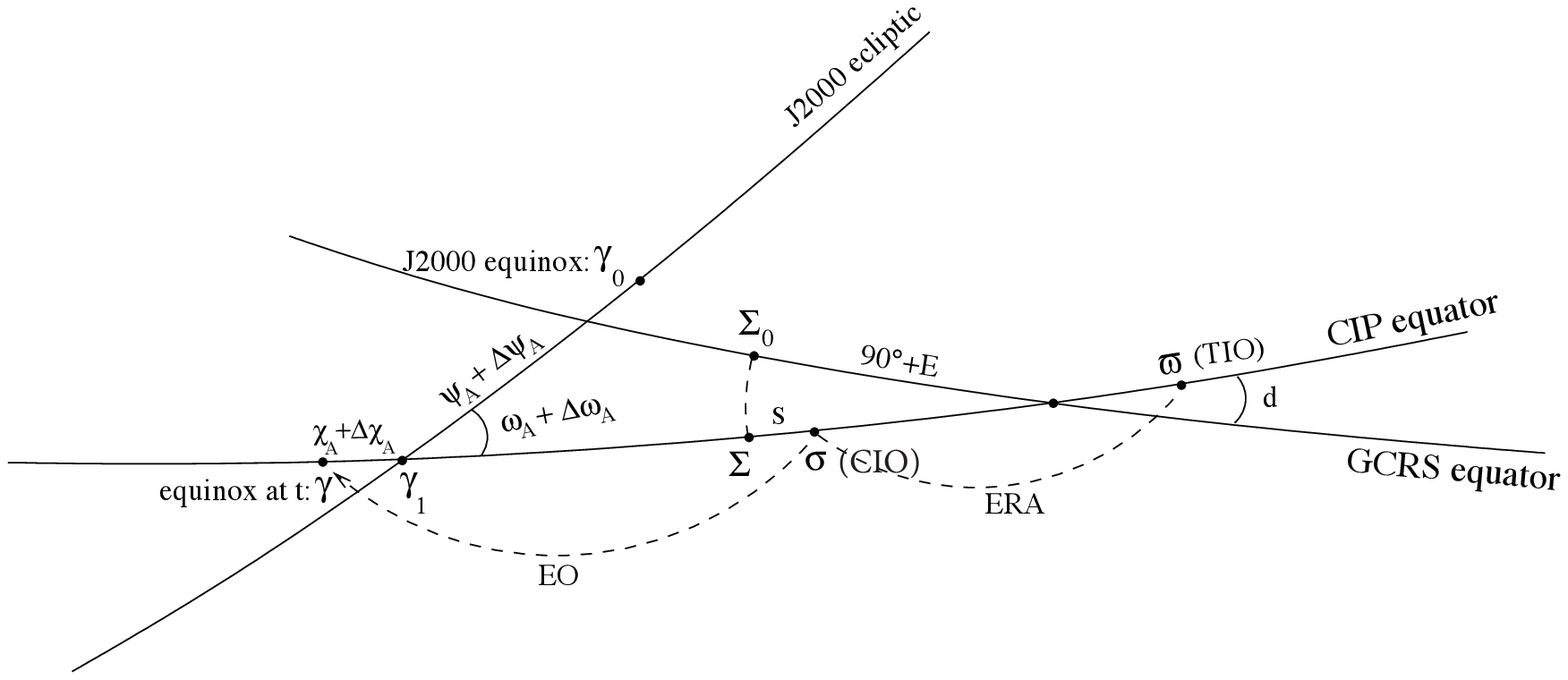} }
\caption{{\small {\bf left} - Orientation of the CIP unit vector: $E$ and $d$ are the GCRS polar coordinates (with $X = \sin d \cos E$; $Y = \sin d \sin E$); $F$ and $g$ are the ITRS polar coordinates (with $x = \sin g \cos F$, $y = \sin g \sin F$); {\bf right} - ERA is the Earth Rotation Angle along the CIP equator; EO is the ``equation of the origins'' that links the CIO and the equinox  $\gamma$; $\gamma_0$ and  $\gamma_1$ are the intersections of the J2000 ecliptic (see above) with the J2000 and CIP equators, respectively; $\psi_A$ and $\Delta\psi_A$, and $\omega_A$ and $\Delta\omega_A$ are the precession and nutation quantities in longitude and obliquity referred to the J2000 ecliptic; $\chi_A+\Delta\chi_A$ is for the ecliptic motion along the CIP equator.}}
\label{figEOP}
\end{center}
\end{figure}

The transformation between the GCRS (the transformed BCRS/ICRS) and the ITRS depends on Earth's rotation, that can be represented by the time-dependent Earth orientation parameters (EOP), for precession-nutation, polar motion and the rotation angle. The IAU~2000 and IAU~2006 resolutions provided accurate definitions for the pole (the Celestial intermediate pole, CIP) and for new origins on the equator (the Celestial and Terrestrial intermediate origins, CIO and TIO) defining the EOP (see Fig.~\ref{figEOP}) as well as the celestial and terrestrial intermediate reference systems (CIRS and TIRS, respectively). These concepts and definitions do not require the use of an ecliptic. 
\vspace{-0.5cm}

In that context, right ascension and declination are considered as being generic terms which can refer to any equator and any origin on that equator (equinox, CIO, ICRS origin, etc.), with the consequence that the equinox, and therefore the ecliptic, is no longer required for expressing coordinates of celestial objects.
The IAU NFA Working Group (cf.$^1$) has provided a chart explaining the CIO based reduction process from ICRS to ITRS coordinates of the directions of stars that specifies the successive celestial reference systems, i.e. ICRS, BCRS, GCRS, CIRS, TIRS and ITRS,  to which the coordinates are referred and the time scale to use. This does not require the use of an ecliptic.
\vspace*{0.7cm}

\noindent {\large 3. PRECESSION-NUTATION AND SOLAR SYSTEM EPHEMERIDES}

\noindent {\it IAU Precession-nutation}

The current IAU precession-nutation model is composed of the IAU~2000A nutation and the IAU~2006 precession. The IAU~2000 nutation, denoted MHB2000, was obtained by Mathews et~al.~(2002) from the REN2000 rigid Earth nutation series of  Souchay et~al.~(1999) for the axis of figure and the MHB2000 ``transfer function'' to transform from rigid to non-rigid Earth. The series for nutation in longitude and obliquity with their time variations, include 1365 terms of luni-solar and planetary origins, with ``in-phase'' and ``out-of-phase'' components and arguments that are functions of the fundamental arguments of the nutation theory, all these parameters referring to a time-dependent (i.e. moving) ecliptic. 

The IAU~2006 precession (Capitaine et al.~2003) provides polynomial expressions up to the 5th degree in time $t$, both for the precession of the ecliptic and the precession of the equator. The precession of the ecliptic (i.e. the $P_A$ and $Q_A$ parameters with respect to the ecliptic and equinox of J2000.0) was computed as the part of the motion of the ecliptic covering periods longer than 300 centuries, while shorter ones are presumed to be included in the periodic component of the ecliptic motion (VSOP87 + fit to DE406). The precession of the equator was derived from the dynamical equations expressing the motion of the mean pole about the ecliptic pole. The solution includes the geodesic precession due to the relativistic rotation of the ``dynamically non-rotating'' geocentric frame in which the precession equations are solved with respect to the GCRS in which precession-nutation is actually observed. The convention for separating precession from nutation, as well as the integration constants used in solving the equations, have been chosen in order to be consistent with the IAU~2000A nutation.  For continuity reasons, the choice of precession parameters has been left to the user. Therefore, the IAU precession for the equator provides polynomial developments for a number of quantities for use in both the equinox based and CIO based paradigms. The series for the $X$, $Y$ GCRS CIP coordinates (cf.~Fig.~1) consistent with the IAU~2006 precession and IAU~2000 nutation are provided in the IERS Conventions (2010).
 \smallskip
 
\noindent {\it Definition of the ecliptic}

IAU~2006 Resolution~B1 that adopted the new precession, also clarified some aspects of the definition of the ecliptic. First, it recommended that the terms {\it lunisolar precession} and {\it planetary precession} be replaced by {\it precession of the equator} and {\it precession of the ecliptic}, respectively, in order to make clear that they are due to different physical phenomena. Second, it recommended that the ecliptic pole be explicitly defined by the mean orbital angular momentum vector of the Earth-Moon barycenter (EMB) in the Barycentric Celestial Reference System (BCRS). However, even with that improvement, the concept of an ecliptic is not as clear as those of the other astronomical references introduced by the IAU~2000-2009 resolutions (cf.~Sect.2), e.g. for defining the ``mean'' of the  orbital angular momentum vector, or, as already noted in Sect.~2, for defining the GCRS ecliptic.
\smallskip

\noindent {\it The use of the ecliptic in the theory of precession-nutation}

Semi-analytical solutions for the precession-nutation of the CIP equator in the GCRS can be obtained by solving the differential equations for the Euler angles $\psi_A$, and $\omega_A$ (cf. Bretagnon et al.~1997), or for the $X$ and $Y$ GCRS CIP coordinates (cf. Capitaine et al.~2005), given the semi-analytical developments of the luni-solar and planetary torques acting on the oblate Earth and integration constants; those are estimated by VLBI observations, which are only sensitive to the equator and insensitive to the ecliptic. Such solutions do not need the use of a time-dependent (i.e. moving) ecliptic.

However, the IAU~2006 precession computations followed the traditional way in which a time-dependent ``ecliptic" (i.e. a kind of GCRS representation of the mean EMB orbital motion) was used as an intermediate plane for expressing the contributions to the precession rates\footnote[2]{Note that the semi-analytical expressions of these rates referring to this intermediate frame are reduced as compared to expressions referred to the fixed ecliptic.} resulting from the external torque. The values that were the best compatible with the IAU~2000 nutation, were, at the time of the computation, the components in longitude and obliquity respectively, expressed in an equatorial frame linked to the moving equinox. Therefore, the integration of the equations with respect to a fixed ecliptic, required to apply a rotation by the angle $\chi_A$ that expresses the precession of the ecliptic (see Fig.~1).  
Such an approach, although being still quite correct at a certain level of accuracy, should be considered as belonging to the past, i.e. to a transition step where there was the need of providing consistent developments for all the precession parameters, some of them mixing equator and ecliptic for continuity reasons. It should be clear that the use of an ecliptic will not be required for providing a future semi-analytical precession-nutation solution.  The precession of the ecliptic is not necessary any longer.
 
\smallskip

\noindent {\it Modern Solar System ephemerides}

Solar system ephemerides provide the positions and motions of the major planetary bodies in the solar system, including the Earth, Moon and Sun, to very high precision. The 3 state-of-the-art numerical solar system ephemerides, namely the American one, DE (Development Ephemeris; JPL), the Russian one, EPM (Ephemerides of Planets and the Moon; IPA, St.Petersburg) and the French one, INPOP (Int\'egrateur Num\'erique Plan\'etaire de l'Observatoire de Paris) are based on the post-Newtonian equations of motion for a set of "point-masses", the Einstein-Infeld-Hoffmann (EIH) equations and the consideration of various dynamical effects such as figure effects, Earth tides, Earth rotation, lunar librations and tidal dissipation (Soffel \& Langhans 2013). These equations are integrated numerically for the whole solar system including a set of selected minor planets. While the DE100-series ephemeris and the DE200 one were in the B1950 and the J2000 coordinate systems respectively (which referred to the equinox of epoch), the DE400 series (Folkner et al. 2014), as well as the recent EPM (Pitjeva et al. 2013) and INPOP (Fienga et al. 2011) solutions, are oriented to the ICRF system using ICRF-based VLBI measurements of spacecrafts near planets; therefore, theses ephemerides can be considered as being dynamical realizations of the ICRF without the need of any ecliptic. 
\vspace*{0.7cm}

\noindent {\large 4. CONCLUDING REMARKS}
\smallskip

\noindent {\it The role of the ecliptic in modern astronomy}

We can conclude from the previous sections that (i)~no ecliptic is needed for the realization of the reference systems currently used in astronomy, (ii)~the ecliptic is no more needed as a reference for the astronomical coordinates, (iii)~the modern numerical barycentric ephemerides are referred to the ICRF, (iv)~the modern description of precession-nutation of the equator is the motion of the CIP in the GCRS without reference to the ecliptic, (v)~numerical integration, as well as modern semi-analytical integration, of precession-nutation do not use an ecliptic. However, for continuity with the traditional approach, it may be useful to define a conventional BCRS fixed ecliptic frame as realized by rotating the BCRS by a constant rotation according to some mean ecliptic and equinox J2000; note that additional conventions would then be necessary for defining its GCRS counterpart.

Therefore, while the equinox (and the tropical year) will always have some value for the seasons, the organisation of everyday life and the calendar,  the geometric, kinematical and dynamical uses of ecliptic in modern astronomy are now limited to uses for continuity with historical references and parameters. 
\smallskip

\noindent {\it Is a definition of the ecliptic in the framework of GR necessary?}

In relativity, it is necessary to carefully distinguish between barycentric and geocentric quantities, so the calculation of a moving ecliptic presents a serious problem when it is used in the GCRS.  Due to the loss of the importance of the ecliptic, the definition of the time-dependent ecliptic in GR is not required.

\vspace*{0.7cm}

\noindent {\large 5. REFERENCES}
{

\leftskip=5mm
\parindent=-5mm
\smallskip

Bretagnon, P., Rocher, P., Simon, J.-L., 1997, A\&A 319, 305

Capitaine, N., Wallace, P.T., \& Chapront, J., 2003, A\&A 412, 567

Capitaine, N., Folgueira, M. and Souchay, J., 2005,  A\&A 445, 347

Capitaine N. \& the IAU NFA WG, 2007, IAU Transactions 26B, 3, K. van der Hucht~(ed), 14, pp.~74--75

Fienga A., Laskar J. Kuchynka P. et al., 2011, \cmda 111, 363

Folkner et al. 2009,  The planetary and lunar ephemerides DE421, IPN Progress Report 42-178, 1-34

IERS Conventions (2010), IERS Technical Note 36, G. Petit and B. Luzum~(eds), Frankfurt am Main: Verlag des Bundesamts f\"ur Kartographie und Geod\"asie, 2010

Mathews P.M., Herring, T., Buffett, B.A., 2002, \jgr 107, B4, 2068, 10.1029/ 2001JB000390

Pitjeva et al. 2013, \mnras 432, 3431

Soffel, M., Klioner, S. A., Petit, G. et al., 2003, AJ 126, 6, 2687

Soffel, M. \& Langhans R. 2013, ``Space-Time Reference Systems'', Springer-Verlag Berlin Heidelberg

Souchay, J., Loysel B., Kinoshita, H., Folgueira, M.,1999, A\&A Supp. Ser. 135, 111

}

\end{document}